\documentclass[letterpaper]{article} 
\usepackage{aaai23}  
\usepackage{times}  
\usepackage{helvet}  
\usepackage{courier}  
\usepackage[hyphens]{url}  
\usepackage{graphicx} 
\usepackage{natbib}  
\usepackage{caption} 
\usepackage{algorithm}
\usepackage{algorithmic}

\usepackage{newfloat}
\usepackage{listings}

\usepackage{color}
\usepackage{multicol}
\usepackage{array}
\usepackage{stfloats}
\usepackage{amsthm} 
\usepackage{multirow}
\usepackage{amssymb}
\usepackage{float}
\usepackage{nccmath}
\usepackage{graphicx}
\usepackage{setspace}
\usepackage[algo2e, linesnumbered, ruled, vlined]{algorithm2e}
\usepackage{mathtools}
\usepackage{booktabs}
\newcommand{\thetab}{\ensuremath{\bm{\theta}}}
\usepackage{mwe}
\usepackage{cite}
\usepackage{amsmath,amssymb,amsfonts}
\usepackage{graphicx}
\usepackage{textcomp}
\usepackage{xcolor}
\usepackage{physics}
\usepackage{bm}
\usepackage[utf8]{inputenc} 
\usepackage{kotex} 
\usepackage[linesnumbered,ruled]{algorithm2e}
\usepackage{cite}
\usepackage{amsmath,amssymb,amsfonts}
\usepackage{algorithmic}
\usepackage{graphicx}
\usepackage{textcomp}
\usepackage{xcolor}

\newcommand{\lcal}{\ensuremath{\mathcal{L}}}

\newcommand{\BfPara}[1]{{\noindent\bf#1.}\xspace}
\usepackage[normalem]{ulem}

\DeclareCaptionStyle{ruled}{labelfont=normalfont,labelsep=colon,strut=off} 
\floatstyle{ruled}
\newfloat{listing}{tb}{lst}{}
\floatname{listing}{Listing}
\usepackage{multirow}
\usepackage{multicol}
\usepackage{color}
\usepackage{array}
\usepackage{stfloats}
\usepackage{float}
\usepackage{bbm}
\usepackage{nccmath}
\usepackage{graphicx}
\usepackage[normalem]{ulem}
\usepackage{booktabs}
\usepackage{mwe}
\usepackage{cite}
\usepackage{graphicx}
\usepackage[export]{adjustbox}
\usepackage{amsmath,amssymb,amsfonts}
\usepackage{textcomp}
\usepackage{xcolor}
\usepackage[utf8]{inputenc} 
\usepackage{kotex} 

\usepackage{grffile}
\usepackage{xurl}

\title{FV-Train: Quantum Convolutional Neural Network Training with a Finite Number of Qubits by Extracting Diverse Features}
\author{
    Hankyul Baek\textsuperscript{\rm 1}, Won Joon Yun\textsuperscript{\rm 1}, Joongheon Kim\textsuperscript{\rm 1}\thanks{Corresponding author}
}
\affiliations{
    \textsuperscript{\rm 1}Korea University, Seoul, Republic of Korea\\
    \{67back,ywjoon95,joongheon\}@korea.ac.kr
}

\usepackage{bibentry}

\begin{document}

\maketitle

\begin{abstract}
Quantum convolutional neural network (QCNN) has just become as an emerging research topic as we experience the noisy intermediate-scale quantum (NISQ) era and beyond. As convolutional filters in QCNN extract intrinsic feature using quantum-based ansatz, it should use only finite number of qubits to prevent barren plateaus, and it introduces the lack of the feature information. In this paper, we propose a novel QCNN training algorithm to optimize feature extraction while using only a finite number of qubits, which is called fidelity-variation training (FV-Training). 
\end{abstract}

\section{Introduction}\label{sec:introduction}
Quantum computing has recently received the spotlight because of its potential to solve complex problems faster than classical algorithms. In contrast to classical computation, which uses a linear scale in bits, quantum computing uses an exponential scale in qubits. It's because the entanglement of qubits can allow to represent multiple states simultaneously. As a result, quantum machine learning (QML) has attained linear or sublinear complexity in comparison to the polynomial complexity of conventional machine learning, even in the current era of noisy intermediate scale quantum (NISQ). Accordingly, various research utilize QML to optimize its objectives~\cite{aimlab2022icte,aimlab2022dynn}. However, there is still a challenging problem in utilizing QML~\textit{i.e.}, barren plateaus. Barren plateaus hinder the training of QML, and a lot of research has proved that increasing qubits in ansatz induces barren plateaus~\cite{mcclean2018barren}. In this paper, we aim to optimize QML, especially quantum-based CNN (QCNN), using only a finite number of qubits to prevent barren plateaus while maintaining reasonable performance. This approach is called fidelity-variation training (FV-Train) in this paper. The novelty of our proposed FV-Train is numerically and experimentally evaluated, and we finally confirm our proposed method achieves desired performance improvements.

\section{Related Work}

\BfPara{Basic Quantum Gates}A qubit is a two-state quantum-mechanical computing unit where the quantum state is represented with two basis states $|0\rangle ,|1\rangle$ in Bloch sphere~\cite{guan2021quantum}. The quantum state can be described as $|\psi\rangle= \alpha|0\rangle + \beta|1\rangle$, where $\alpha^2+\beta^2=1$. To utilize a single quantum computing system, a classical data $\delta$ is embedded to quantum state with the rotation gates $R_x(\delta)$, $R_y(\delta)$, and $R_z(\delta)$, where each gate represents the rotation of $\delta$ over $x$-, $y$-, and $z$-axes in Bloch sphere, respectively. By entangling qubits with \textit{controlled-NOT} (CNOT) gates in a multi-qubit system, quantum computing achieves an advantage in processing speed. 

\BfPara{Quantum CNN (QCNN)}\label{sec:quanvolution}
QCNN, also called quanvolutional neural network, is a quantum version of CNN for 2D images using quantum circuits as a convolutional filter~\cite{DBLP:journals/qmi/HendersonSPC20}. To fully leverage massively parallel computations on the superposition of quantum states with finite number of qubits, QCNN designs random quantum circuit layers to extract the intrinsic features of target images. The main challenge in the research of QCNN is minimizing the number of qubits while the ansatz-based filters carry out convolution and full intrinsic features extraction.

\begin{figure}[t!]
\centering
\includegraphics[width=\linewidth]{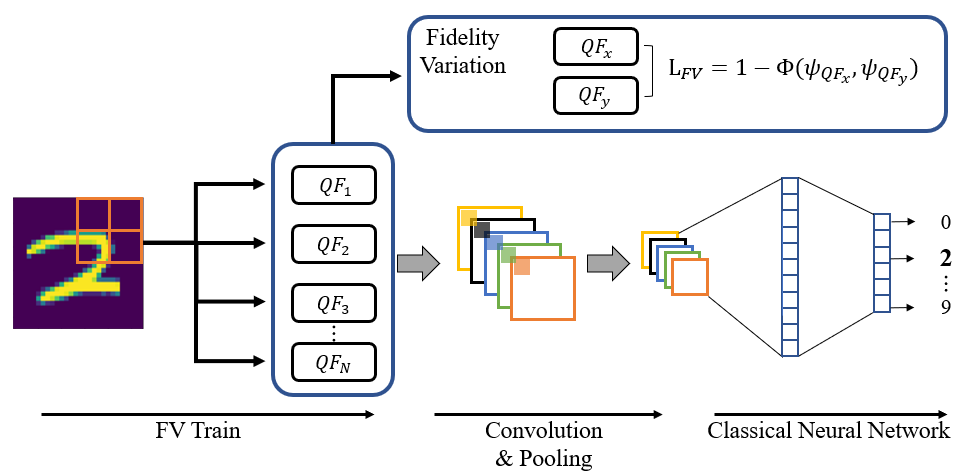}
\caption{Overall Process of FV-Train.}\label{fig:data_reuploading}
\end{figure}

\section{FV-Train}

This section presents our proposed QCNN training algorithm named fidelity-variation training. As mentioned above, using only a small number of qubits while maintaining performance is challenging in QCNN. Note that in QCNN, each ansatz acts like a convolutional filter in classical CNN. To extract various features of target image, we aim to vary the nature of the quanvolutional filter. For this purpose, we design an FV regularizer $\mathcal{L_{FV}}$, motivated by Uhlmann's fidelity function~\cite{jozsa1994fidelity}. The fidelity of output quantum states from two quanvolutional filters $q_X$ and $q_{Y}$ is defined as $\Phi(\rho_{q_{X}}, \rho_{q_{Y}}) = |\langle \psi_{q_X}|\psi_{q_{Y}}\rangle|^2$, where $\rho_{q_X} =|\psi_{q_X} \rangle \langle \psi_{q_X}|$ and $\rho_{q_{Y}} =|\psi_{q_{Y}} \rangle \langle \psi_{q_{Y}}|$.  As the similarity between the two states increases, the fidelity converges to 1. On the other hand, as the similarity between the two states decreases, the fidelity converges to 0, which means that $q_{Y}$ does not follow $q_X$.
We assume that a reduction in the fidelity between output states of the quanvolutional filters enables the extraction of various intrinsic features. With the assumption, we define FV regularizer as 
\begin{equation}
    \mathcal{L_{FV}} = 1 - {1 \over {}_{L}C_2} \sum_{l,l^{'} \in L} \nolimits \Phi(\psi_{q_{l}}, \psi_{q_{l^{'}}}).
\end{equation}
The training process of QCNN with FV regularizer is described in Algorithm 1. The parameters ($x$, $y$) are denoted as the input data and label, respectively. The predicted label can be produced by activating fully-connected layer on the concatenation of the observable in each filter. Cross-entropy loss is described as $\mathcal{L_{CE}} = -{1\over C}\sum^C_{c=1} \log{p(y_{pred} = y_c|x)}$, where $C$ represents the number of classes. Consequently, the total loss of QCNN can be denoted as,
\begin{equation}
    \mathcal{L}_{total} ={1\over D} \sum_{(x, y) \in \zeta^k} \nolimits [\mathcal{L_{CE}} +\lambda \mathcal{L_{FV}}],
\end{equation}
where $D$ is the batch size and  $\lambda$ is the hyper-parameter of FV regularizer.

\begin{algorithm2e}[t]\label{alg: FV-train}
    \SetCustomAlgoRuledWidth{0.44\textwidth}  
\caption{Fidelity Variation Train (FV-Train)}\label{alg:Training algorithm} 
\textbf{Initialization.} \texttt{QCNN} parameters, $w$;

 \For{ $e = \{1,2,\dots, E\}$}
 {
     \For{ $(x, y) \in \zeta^k$}
     {
         \For{$l, l^{'} \in \{1,2,\dots,L-1\}$}
         {  
            Get features with $l$-th and $l^{'}$-th filter\;
            Calculate $\mathcal{L_{FV}}$\;
            Calculate loss gradients\;
            
         }
         Calculate $\mathcal{L}_e^k \leftarrow \mathcal{L}_{total}$\;
         $\thetab^k_{e+1}\leftarrow \thetab^k_{e}-\eta_e\nabla_{\theta^k_{e}}\lcal^{k}_{e}$\;
     }    
 }
\end{algorithm2e}

\section{Experiments}

\BfPara{Setting}
To corroborate the performance of the FV-Train, we train a QCNN to classify the MNIST dataset through a classical training method (Vanilla-Train) and the FV-Train with various fidelity regularizer parameter ($\lambda$ = 0.1, 0.5, 1). We design two random ansatz-based convolutional filters to extract the intrinsic features. The initial fidelity between the two filters is set to $0.611$ in all experiment setting.

\begin{figure}[t]
    \centering
    \begin{tabular}{@{}c c@{}}
    \includegraphics[width=.485\columnwidth]{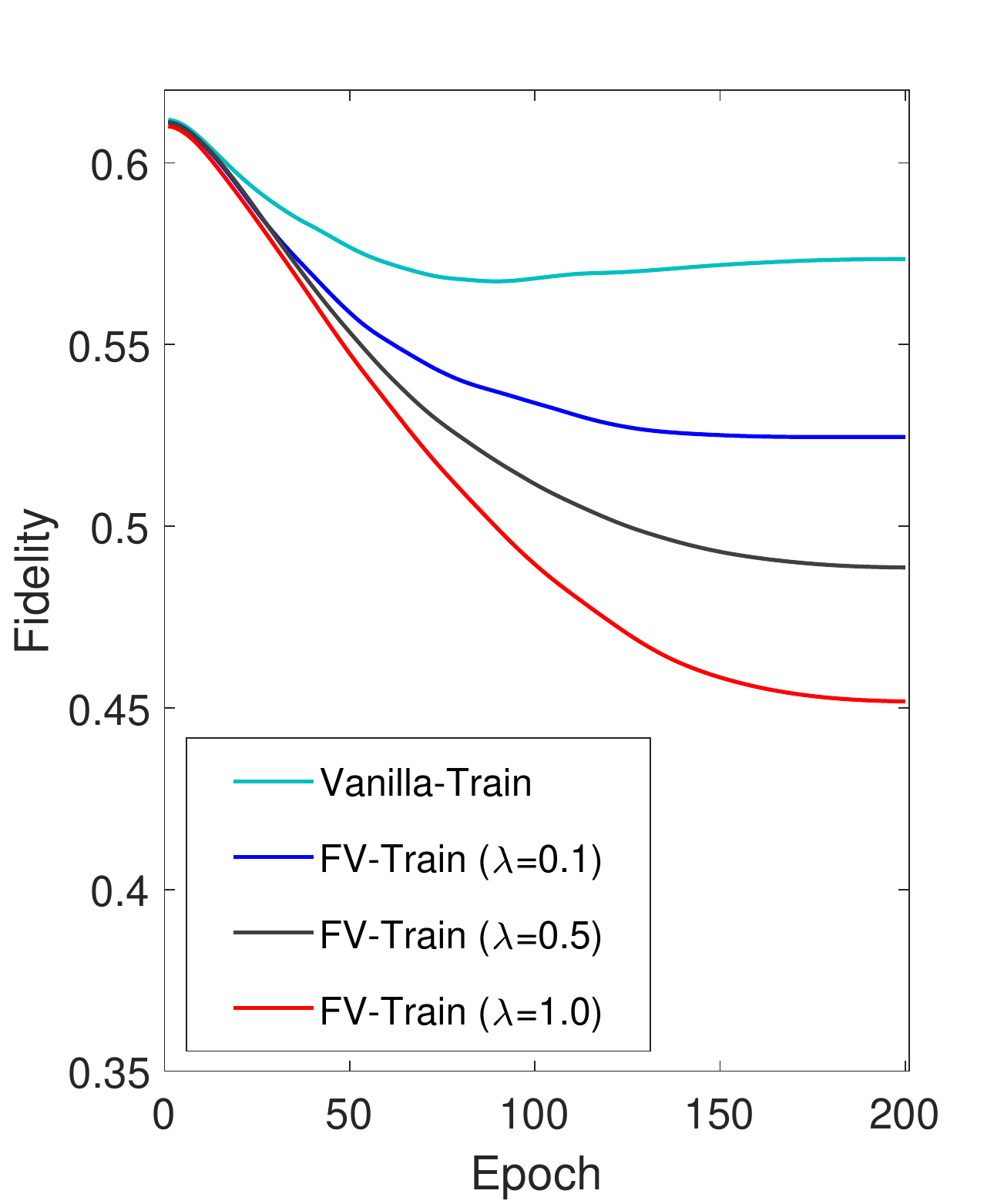}
    &\includegraphics[width=.485\columnwidth]{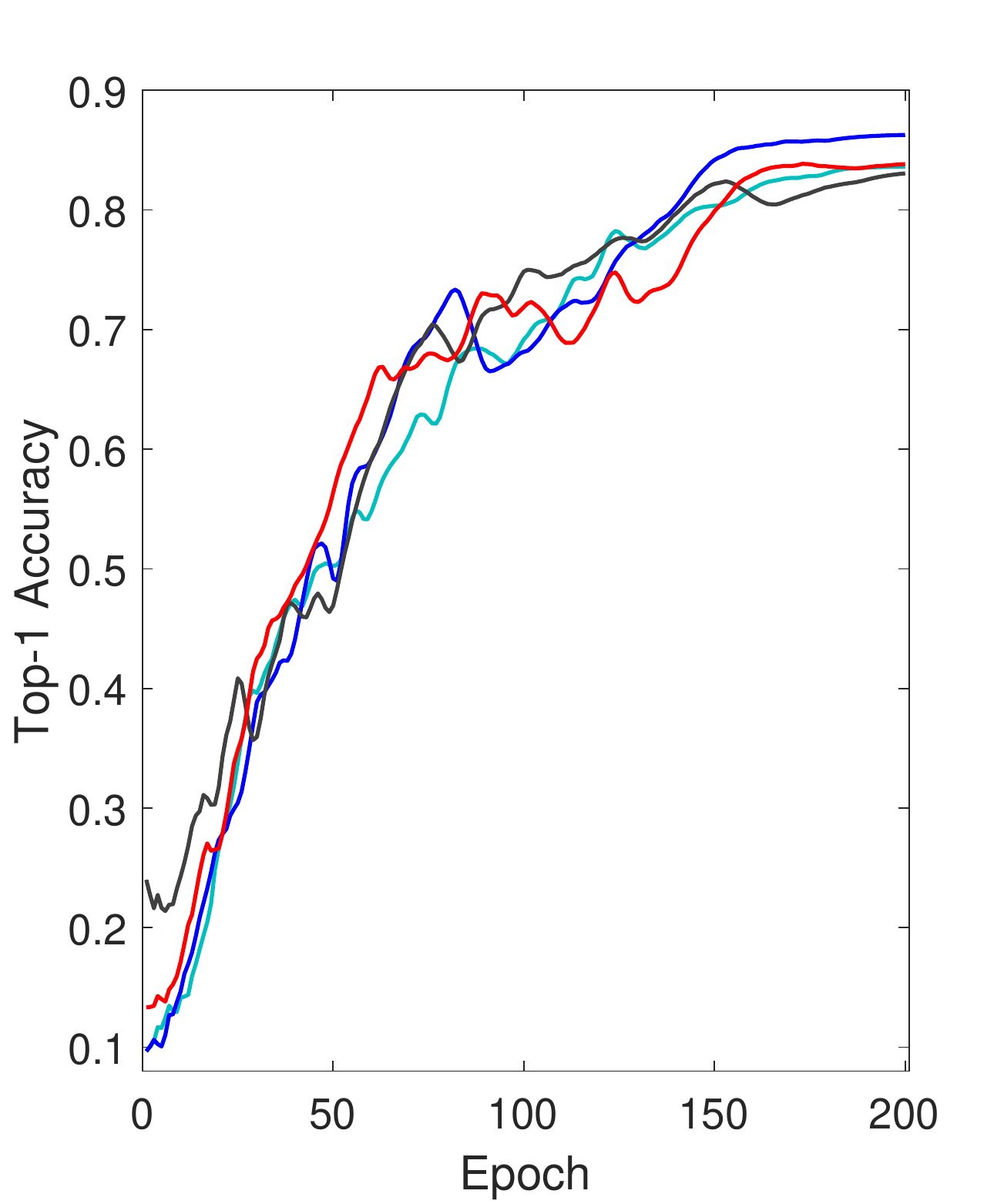}\\
    ~~~~\small (a) Top-1 accuracy. 
    &~~~~~\small (b) Fidelity.
    
    \end{tabular}\vspace{-3mm}
    \caption{Performance evaluation for FV-Train.}
    \vspace{-1mm}
    \label{exp:main}
\end{figure}

\BfPara{Experimental Results}
Fig.~\ref{exp:main} represents the performance difference between the Vanilla-Train and the FV-Train. Fig.~\ref{exp:main} (a) shows the fidelity of each training method. Note that fidelity is a measure of closeness of two quantum states. The FV-Train with different $\lambda$ shows that the features extracted through the FV-Trained filter are more diversified than the features from Vanilla-Trained filter. Here, we confirm our assumption that diversified features results in good performance. In Fig.~\ref{exp:main} (b), we observe that the FV-Train outperforms the Vanilla-Train. The FV-Train ($\lambda=0.1$) achieves 5.5\% higher top-1 accuracy than the Vanilla-Train, and the FV-Train ($\lambda=0.5, 10$) achieves slightly higher accuracy than the vanilla-train. From the results, we confirm that even with the same QCNN, the FV-Train obtains more diverse features than the vanilla-train.

\section{Conclusion}
In this paper, we propose a novel QCNN training framework based on the concept of CNN and QML. To get various filters with finite number of qubits, we design FV-Train regularizer. To corroborate the performance of FV-Train with QCNN, we compare the performance of QCNN with FV-Train and vanilla-train. In future research, we intend to solve the barren plateaus problem of QML by extracting various intrinsic features through FV-Train while using only a finite number of qubits.

\small
\bibliography{PTCL}

\begin{thebibliography}{6}
\providecommand{\natexlab}[1]{#1}

\bibitem[{Guan et~al.(2021)Guan, Perdue, Pesah, Schuld, Terashi, Vallecorsa,
  and Vlimant}]{guan2021quantum}
Guan, W.; Perdue, G.; Pesah, A.; Schuld, M.; Terashi, K.; Vallecorsa, S.; and
  Vlimant, J.-R. 2021.
\newblock Quantum machine learning in high energy physics.
\newblock \emph{Machine Learning: Science and Technology}, 2(1): 011003.

\bibitem[{Henderson et~al.(2020)Henderson, Shakya, Pradhan, and
  Cook}]{DBLP:journals/qmi/HendersonSPC20}
Henderson, M.; Shakya, S.; Pradhan, S.; and Cook, T. 2020.
\newblock Quanvolutional Neural Networks: Powering Image Recognition with
  Quantum Circuits.
\newblock \emph{Quantum Machine Intelligence}, 2(1): 1--9.

\bibitem[{Jarrod R.~McClean et~al.(2018)Jarrod R.~McClean, Smelyanskiy,
  Babbush, and Neven}]{mcclean2018barren}
Jarrod R.~McClean, S.~B.; Smelyanskiy, V.~N.; Babbush, R.; and Neven, H. 2018.
\newblock Barren plateaus in quantum neural network training landscapes.
\newblock \emph{CoRR}, abs/1803.11173.

\bibitem[{Jozsa(1994)}]{jozsa1994fidelity}
Jozsa, R. 1994.
\newblock Fidelity for mixed quantum states.
\newblock \emph{Taylor \& Francis Journal of Modern Optics}, 41(12):
  2315--2323.

\bibitem[{Kwak et~al.(2022)Kwak, Yun, Kim, Cho, Choi, Jung, and
  Kim}]{aimlab2022icte}
Kwak, Y.; Yun, W.~J.; Kim, J.~P.; Cho, H.; Choi, M.; Jung, S.; and Kim, J.
  2022.
\newblock Quantum Heterogeneous Distributed Deep Learning Architectures:
  Models, Discussions, and Applications.
\newblock \emph{CoRR}, abs/2202.11200.

\bibitem[{Yun et~al.(2022)Yun, Kim, Jung, Park, Bennis, and
  Kim}]{aimlab2022dynn}
Yun, W.~J.; Kim, J.~P.; Jung, S.; Park, J.; Bennis, M.; and Kim, J. 2022.
\newblock Slimmable Quantum Federated Learning.
\newblock In \emph{Proc. of the ICML Workshop on Dynamic Neural Networks}.

\end{thebibliography}

\vspace{1mm}
\BfPara{Acknowledgments}
This research was funded by National Research Foundation of Korea (2022R1A2C2004869).
\end{document}